\documentclass[twocolumn,trackchanges]{aastex7}

\usepackage{textcomp}
\usepackage{xspace}
\usepackage{amsmath}
\usepackage{graphicx}
\usepackage{float}
\usepackage{gensymb}
\usepackage{comment}

\newcommand {\ixpe}{\text{IXPE}\xspace}
\newcommand{\chandra}{\text{Chandra}\xspace}
\newcommand{\xmm}{\text{XMM-Newton}\xspace}
\newcommand{\xrism}{\text{XRISM}\xspace}

\begin{document}

\title{What's the Buzz About GX~13$+$1? Constraining Coronal Geometry with QUEEN-BEE: \\ A Bayesian Nested Sampling Framework for X-ray Polarization Rotation Analysis}

\author[orcid=0000-0002-2381-4184,sname=Ravi,gname=Swati]{Swati Ravi}
\affiliation{MIT Kavli Institute for Astrophysics and Space Research, Massachusetts Institute of Technology \\ 77 Massachusetts Avenue, Cambridge, MA 02139, USA}
\email[show]{swatir@mit.edu}  

\author[orcid=0000-0002-0940-6563, sname=Ng,gname=Mason]{Mason Ng} 
\affiliation{Department of Physics, McGill University \\ 3600 rue University, Montréal, QC H3A 2T8, Canada}
\affiliation{Trottier Space Institute, McGill University \\ 3550 rue University, Montréal, QC H3A 2A7, Canada}
\affiliation{MIT Kavli Institute for Astrophysics and Space Research, Massachusetts Institute of Technology \\ 77 Massachusetts Avenue, Cambridge, MA 02139, USA}
\email{mason.ng@mcgill.ca}

\author[orcid=0000-0002-6492-1293, sname=Marshall,gname=Herman]{Herman L. Marshall}
\affiliation{MIT Kavli Institute for Astrophysics and Space Research, Massachusetts Institute of Technology \\ 77 Massachusetts Avenue, Cambridge, MA 02139, USA}
\email{hermanm@mit.edu}

\author[0000-0002-0642-1135, sname=Gnarini,gname=Andrea]{Andrea Gnarini}
\affiliation{Dipartimento di Matematica e Fisica, Universit\`{a} degli Studi Roma Tre\\ Via della Vasca Navale 84, I-00146 Roma, Italy}
\email{andrea.gnarini@uniroma3.it}

\begin{abstract}

Observations from the Imaging X-ray Polarimetry Explorer (\ixpe) have revealed electric vector position angle (EVPA) rotation in several neutron star low-mass X-ray binaries, including the galactic X-ray burster GX~13$+$1. We developed a novel Bayesian nested sampling framework---``Q-U Event-by-Event Nested sampling for Bayesian EVPA Evolution'' (QUEEN-BEE)---to model unbinned Stokes parameters and infer optimal EVPA rotation rates in \ixpe data. We then applied this framework to three previous \ixpe observations of GX~13$+$1. In the first observation, QUEEN-BEE recovers a rotation rate of $42\pm4^\circ/\mathrm{day}$, consistent with prior binned analysis. Energy-binned QUEEN-BEE analysis of this first observation suggests a slab-like coronal geometry, providing the first constraints between slab and shell coronae for this source. We also explore alternative EVPA rotation scenarios in GX~13$+$1 including variable disk wind behavior. The second observation of this source shows no evidence of rotation, and the third observation shows transient rotating behavior with an EVPA rotation rate when exiting a light curve dip of $170^{+20}_{-40}\,^\circ/\mathrm{day}$. The results show marginal but consistent increases in the overall measured polarization degree (PD) for epochs where the EVPA rotation is identified. These results demonstrate that QUEEN-BEE can identify evolving polarization signatures in both time- and energy-resolved regimes, even where binned methods fall below detection thresholds. Our findings highlight the diagnostic potential of QUEEN-BEE as a tool for discriminating between competing physical models of coronal geometry and probing disk-wind-related polarization behavior, highlighting the promising potential for application of this framework in a variety of other \ixpe observations.

\end{abstract}

\keywords{\uat{Polarimetry}{1278} --- \uat{Bayesian Statistics}{1900} --- \uat{Neutron Stars}{1108} --- \uat{X-ray binary stars}{1811} --- \uat{Accretion}{14} --- \uat{X-ray Astronomy}{1810} --- \uat{Astrostatistics}{1882}}


\section{Introduction} \label{sec:intro}
\subsection{X-ray Polarimetry}
The 2021 launch of the NASA-ASI Imaging X-ray Polarimetry Explorer (\ixpe) marked the beginning of a new era in X-ray polarimetry, offering the first opportunity in decades to systematically probe the high-energy universe through spatially, temporally, and spectrally resolved polarization measurements \citep{2022JATIS...8b6002W}. The instrument, designed to measure the linear polarization of X-ray emission in the 2--8~keV energy range, consists of three identical telescopes, each comprising a Wolter-I grazing-incidence mirror assembly paired with a gas-pixel detector (GPD). Incoming X-rays focused by the mirrors interact with the GPDs, producing photoelectrons whose emission directions are modulated according to the polarization of the incident photons, with the distribution of these directions in azimuthal angle space revealing both the polarization degree (PD) and polarization angle (PA) for each detected event \citep{2021AJ....162..208S}. Since its launch, \ixpe has observed over 100 unique targets ranging from galactic sources such as black hole and neutron star X-ray binary systems to extragalactic active galactic nuclei (AGN), placing constraints on coronal geometries and jet magnetic field structures of these sources \citep[e.g.,][]{2023MNRAS.523.4468G, 2023NatAs...7.1245D}.

\subsection{X-ray Polarimetric Timing}
An emerging frontier in X-ray polarimetric analysis is polarimetric timing---the study of short-term evolution in the electric vector position angle (EVPA) over a single observation (see \cite{2022hxga.book..113I} for a detailed overview.) Such temporal evolution of the EVPA within an observation can provide insights into dynamic accretion processes such as disk or jet precession and variable scattering phenomena arising from warped or tilted disks \citep{2022NatAs...6.1433D, 2023arXiv231103667H}. Furthermore, polarimetric timing is essential to accurately constrain the true polarization signal of a system, since the average PD may deviate from the intrinsic PD if there is strong variability in the EVPA. Without accounting for this variability, even strong intrinsic polarization can appear weak or absent, leading to incorrect inferences about the source behavior. 

A diverse collection of sources previously observed by \ixpe have revealed instances of evolving EVPAs. Some sources like the blazar Mrk~421 \citep{2023NatAs...7.1245D} and the pulsar 4U~1626$-$67 \citep{2022ApJ...940...70M} exhibited statistically insignificant time-averaged PDs, but significant intrinsic PDs with subsequent time- and phase-binned analyses. \ixpe observations of Mrk 421 specifically revealed smooth EVPA rotation of more than $360^\circ$ in five days, consistent with optical polarimetry measurements \citep{2023NatAs...7.1245D}. Rotational behavior in EVPA has also been observed in several neutron star low-mass X-ray binaries (NS-LMXBs)---gravitationally bound systems including both a neutron star and low-mass companion ($\lesssim$ 1 \(M_\odot\)) \citep{2025A&A...699A.230G}. Sco~X-1, Cir X-1, and GX~13$+$1, for example, have revealed EVPA rotations over timescales ranging from days to decades \citep{2024ApJ...960L..11L, 2024ApJ...961L...8R, 2024A&A...688A.170B}. Sco~X-1 exhibited a rotation of up to ~$45\pm8^\circ$ between observations taken in 1977 and 2023 taken by the 8\textsuperscript{th} Orbiting Solar Observatory (OSO-8) \citep{1979ApJ...232L.107L} and \ixpe respectively, moving from a PA aligned parallel with the jet direction to nearly perpendicular alignment with the jet \citep{2024ApJ...960L..11L}. On shorter timescales, Cir X-1 exhibited a rotation in the polarization angle by $49\pm8^\circ$ between two segments of \ixpe observation separated by six days, possibly attributed to misalignment of spin and orbital axes \citep{2024ApJ...961L...8R}. More dramatically, the first \ixpe observation of GX~13$+$1 directly observed an EVPA rotation of $\sim70^\circ$ within a single 2-day observation, with subsequent \ixpe observations of the source suggesting further EVPA rotation \citep{2024A&A...688A.170B, 2025ApJ...979L..47D, 2025arXiv250805763K}.

\subsection{GX~13$+$1} 
Among the sources with observed EVPA  rotation, GX~13$+$1 stands out as an enigmatic NS-LMXB, exhibiting massive variability in its luminosity along with correlated timing and spectral properties. It moves between lower luminosity states (traditionally the ``atoll state") and luminosities close to Eddington luminosity (the ``Z-state"), thought to be driven by a varying mass accretion rate \citep{1989A&A...225...79H, 2015ApJ...809...52F, 2024ApJ...966..232N}.  The source also displays strong quasi-periodic oscillations (QPOs), periodic absorption dips, and an orbital period of approximately 24.5 days \citep{2010ApJ...719..979C, 2014A&A...561A..99I, 2021ASSL..461..263M}.

A unique and widely studied feature of GX~13$+$1 is its disk winds. GX~13$+$1 exhibits some of the most prominent and persistent disk winds of any NS-LMXB. \chandra and \xmm observations of the source revealed highly ionized and consistently blueshifted absorption lines---clear signatures of an outflowing wind \citep{2001ApJ...556L..87U,2002A&A...385..940S,2014MNRAS.438..145M}. Their persistence across spectral states and orbital phases points to a stable launching mechanism, likely driven by radiation pressure in the outer disk \citep{2004ApJ...609..325U}. 

Current polarimetric analysis of GX~13$+$1 provides complementary insight into the accretion geometry of the source with previously observed variability in the polarimetric properties of the source, particularly as seen in the first and third \ixpe observations of the source, thought to be connected to disk wind behavior through temporary obscuration of the central X-ray source \citep{2025ApJ...979L..47D}. Taken in greater detail, the \ixpe observation history of  GX~13$+$1 suggests unusual and discrepant behavior ranging from a smooth EVPA evolution of $\sim70^\circ$ in the first \ixpe observation \citep{2024A&A...688A.170B} to no apparent rotation in a subsequent observation \citep{2024A&A...688A.217B}, and multiple strong $\sim70^\circ$ polarization swings in the third \ixpe observation associated with dips in the source's X-ray light curve \citep{2025ApJ...979L..47D, 2025arXiv250805763K}. Follow-up analysis of the source using Bayesian inference indicates possible bi-modal variability of the polarization angle of GX~13$+$1 \citep{2025arXiv250404775L}.
 
In this paper, we introduce ``Q-U Event-by-Event Nested sampling for Bayesian EVPA Evolution'' (QUEEN-BEE), a Bayesian nested sampling framework for X-ray polarimetric analysis, suitable for use with \ixpe data. QUEEN-BEE builds upon the methodology of the maximum likelihood approach, providing a flexible and statistically rigorous framework for detecting and characterizing EVPA evolution, particularly in low-count or noise-dominated regimes. 
In Section~\ref{sec:Methodology} we outline the methodology of QUEEN-BEE, detailing its statistical foundations and implementation architecture. Section~\ref{sec:GX~13$+$1} presents a case study applying QUEEN-BEE to three epochs of \ixpe observation of GX~13$+$1 with comparison against previously published results. Section~\ref{sec:discussion} discusses the physical implications of our findings and explores the broader applicability of QUEEN-BEE to other X-ray polarimetric investigations beyond GX~13$+$1. Finally, a conclusion follows in section~\ref{sec:conclusion}.

\section{QUEEN-BEE: Methodology and Implementation}
\label{sec:Methodology}
\subsection{Existing Methods for EVPA Evolution Analysis}

\label{sec:prev methods}
Given the physical interest in EVPA rotations and the newfound opportunity to explore such rotations using \ixpe, a few analysis techniques have emerged to identify the temporal evolution of polarimetric properties in \ixpe observations. The PA rotation for the first observation of GX~13$+$1 \citep{2024A&A...688A.170B}, was discovered by binning the \ixpe observation into 5 equal time segments and running a model-independent polarimetric analysis of each segment using the polarization cube (PCUBE) algorithm.\footnote{The algorithm is available in the widely-used \texttt{ixpeobssim} software tool \citep{2015APh....68...45K, 2022SoftX..1901194B}} The EVPA rotation in Mrk 421, in contrast, was identified using an event-based maximum likelihood method \citep{2023NatAs...7.1245D}. 

Both techniques ultimately rely on how individual photon events encode polarization through their azimuthal scattering angles, quantified by Stokes parameters. Stokes parameters consist of \textit{I,} \textit{Q,} and \textit{U}, where I denotes the total intensity of the signal, and \textit{Q} and \textit{U} represent two orthogonal components of the polarization vector in the detector frame.\footnote{We ignore the Stokes parameter \textit{V} describing circular polarization, which is not measured by \ixpe.} The PCUBE algorithm calculates Stokes parameters \((Q_{i}, U_{i})\) from each photon’s measured azimuthal scattering angle, performing a weighted sum of the parameters over a predefined time or energy bin to obtain the net Stokes parameters \((Q, U)\) for that bin \citep{2015APh....68...45K}. For a time-evolving system, the observation can be divided into consecutive time bins, within which the Stokes parameters are summed using the PCUBE algorithm for each bin.

The event-based maximum likelihood method, in contrast, estimates Stokes parameters by directly modeling the distribution of photoelectron azimuthal angles from individual X-ray events. Rather than binning the data, this approach evaluates the probability of observing the full set of angles in an observation given a particular polarization model and finds the parameters that maximize this likelihood. With this method, EVPA evolution may be more directly probed by introducing a rotation parameter into the likelihood function. This enables a global fit to the data that can capture smooth rotational behavior without the need to segment the observation into discrete bins \citep{2021AJ....162..134M,2021ApJ...907...82M,2024ApJ...964...88M}. Additional Stokes parameter estimation techniques are presented in \cite{2024OJAp....7E..35H}, along with machine learning-based approaches described in \cite{2019NIMPA.94262389K,2021NIMPA.98664740P} and \cite{2023A&A...674A.107C}.

While both PCUBE and event-based maximum likelihood estimation have been instrumental in identifying EVPA evolution in \ixpe observations, both are limited in their ability to fully characterize time-dependent polarization behavior, particularly in low signal-to-noise regimes where binning may dramatically degrade the signal. The PCUBE algorithm relies on dividing the observation into discrete time bins, making the analysis sensitive to binning choices that may potentially obscure EVPA variations in observations. In particular, it is often unclear how to choose the appropriate number or duration of bins to reliably recover a physically meaningful rotation, especially when the timescale of the evolution is not known a priori. The event-based maximum likelihood approach resolves ambiguities with time binning by modeling the distribution of photoelectron azimuthal angles directly and allowing for parametric descriptions of EVPA evolution across the full dataset. However, it lacks a built-in framework for comparing competing models, particularly of differing complexity or functional form, limiting its ability to assess whether a time-varying EVPA model is preferred over a constant one (and an unpolarized one) in a statistically robust manner.

\subsection{Modeling X-ray Polarization in \ixpe Data}

IXPE measures the linear polarization of incoming X-ray photons through the distribution of photoelectron emission angles $\phi_{i}$, which are recorded event-by-event. The modulation angle probability distribution follows
\begin{equation}
f(\phi) = \frac{1}{2\pi} \left[ 1 + \mu p \cos{2(\phi - \psi)} \right],
\label{eq:distribution}
\end{equation}
where \textit{p} is the PD, $\psi$ is the EVPA (equivalently described as ``polarization angle''), $\mu$ is the instrument modulation factor and $\phi$ is the azimuthal angle for each photoelectron event. The joint likelihood across all \textit{N} photon events is thus a product of event-based likelihoods following the form

\begin{equation}
\mathcal{L}(p, \psi) = \prod_{i=1}^{N} f(\phi_i \mid p_{i}, \psi_{i}).
\label{eq: likelihood}
\end{equation}
While EVPA $\psi$ could, in principle, depend on any number of physical variables, here we restrict our attention to time-dependent models where $\psi=\psi(t)$.

\subsection{Three Competing Polarization Models}
\label{subsec:models}

In QUEEN-BEE we test three hypotheses for the EVPA behavior of the source---constant EVPA, linearly rotating EVPA, and unpolarized.

The constant EVPA model, implemented by the \texttt{scout\_const()} function in QUEEN-BEE, assumes the PD \textit{p} and EVPA \textit{$\psi$} are constant over the entire observation. Rather than parametrizing directly in \textit{p} and \textit{$\psi$}, the likelihood is computed in terms of the normalized Stokes parameters (\(q=Q/I, u=U/I\)) through the following standard relationships\footnote{Note that angles are defined as 2\textit{$\psi$} to preserve the $180^\circ$ symmetry in EVPA}

\begin{equation}
q = p \cos(2\psi), 
\label{eq:q}
\end{equation}
\begin{equation}
u = p \sin(2\psi),
\label{u}
\end{equation}
\begin{equation}
p = \sqrt{q^2 + u^2},
\label{eq:p}
\end{equation}
\begin{equation}
\psi = \frac{1}{2} \tan^{-1}(u/q).
\label{eq:psi}
\end{equation}
Substituting these relationships into the likelihood function defined in equations~\ref{eq:distribution} and~\ref{eq: likelihood} therefore becomes\footnote{To avoid an undefined likelihood evaluation in instances when $q=0$, it is useful to substitute $\tan^{-1}\!\big(\frac{u}{q}\big)$ with the equivalent form
$\sin^{-1}\!\big(\frac{u}{\sqrt{u^{2}+q^{2}}}\big)$ in the likelihood function described in equation \ref{eq:likelihood_long}.}

\begin{align}
\mathcal{L}(q, u) &= \prod_{i=1}^{N} \frac{1}{2\pi} \Bigg[ 1 + \mu \sqrt{q^2 + u^2} \, \cos \Big( 2\phi_i \notag \\
&\qquad - \tan^{-1}\left(\frac{u}{q}\right) \Big) \Bigg].
\label{eq:likelihood_long}
\end{align}

In the rotating model, implemented by the \texttt{scout\_rot()} function in QUEEN-BEE, the EVPA is assumed to evolve smoothly over time according to a constant rotation rate $\omega$, such that 

\begin{equation}
\psi(t) = \psi_{0} + \omega t,
\label{eq:psioft}
\end{equation}
where $\psi_{0}$ is the inital EVPA angle, and $\omega$ is the EVPA rotation rate in units of degrees per day. The instantaneous Stokes parameters become 

\begin{equation}
q(t) = p \cos(2\psi_{0}+\omega t), 
\label{eq:q(t)}
\end{equation}
\begin{equation}
u(t) = p\sin(2\psi_{0}+\omega t),
\label{u}
\end{equation}
and the likelihood is evaluated as 

\begin{equation}
\mathcal{L}(p, \psi_0, \omega) = \prod_{i=1}^{N} \frac{1}{2\pi} \left[ 1 + \mu p \cos 2\left(\phi_i - \psi(t_i)\right) \right].
\end{equation}
This model captures systematic rotation of the EVPA across the duration of an observation.

Finally, the null hypothesis of an unpolarized source, implemented by the \texttt{scout\_unpolar()} function in QUEEN-BEE, assumes the source emits unpolarized radiation, i.e., \textit{q = 0}, \textit{u = 0}. In this case, the modulation angle distribution would be uniform, where

\begin{equation}
f(\phi) = \frac{1}{2\pi}.
\label{eq:uniform_distribution}
\end{equation}
This model has no free parameters and serves as a reference point for evaluating preference for any polarization detection through comparing model Bayesian evidences. 

These likelihood formalisms take a simplified, direct, event-based form. For a more general weighted log-likelihood formalism for \ixpe data incorporating detailed background distribution and detector response models appropriate for background-weighted extended sources and instrument simulation contexts, see \cite{2021AJ....162..134M,2021ApJ...907...82M,2024ApJ...964...88M}. These extended features are in development within QUEEN-BEE and will be presented in future work

\subsection{Incorporating Bayesian Nested Sampling for Model Comparison}

QUEEN-BEE applies Bayesian nested sampling to the likelihood formalism defined previously to evaluate how well each of the three different polarization models---constant EVPA, linearly rotating EVPA, and unpolarized---describe the azimuthal angle distribution observed by \ixpe for a given source observation. The aim is to not only infer model parameters (\textit{q}, \textit{u}, \textit{$\omega$}) but also to compute Bayesian evidence for each model to rigorously assess which is favored by the data. 

Nested sampling is particularly suitable for X-ray polarization likelihood analysis, especially of rotating EVPA models, as corresponding likelihood functions can be highly nonlinear, multimodal, and subject to degeneracies between parameters, particularly between PD \textit{p} and PA \textit{$\psi$} (see equations~\ref{eq:p} and~\ref{eq:psi}). Standard MCMC methods can sample the posterior but do not provide a robust framework for model comparison and don't perform well with multimodal likelihoods. Nested sampling, in contrast, provides both posterior samples and model evidence $Z$. 

Nested sampling has already become a broadly adopted tool across astrophysics, including in gravitational-wave parameter estimation and in cosmological model selection, such as tests of inflationary scenarios \citep[e.g.,][]{2009MNRAS.398.1601F, 2011PhRvD..83h2002D, 2015MNRAS.450L..61H, 2020MNRAS.498.4492S, 2021ApJ...908...97L}. Nested sampling, as implemented in QUEEN-BEE using \texttt{dynesty} \citep{2020MNRAS.493.3132S}, works by transforming a multidimensional evidence integral into a one-dimensional integral over the prior mass X defined by

\begin{equation}
    X(\lambda) = \int_{\mathcal{L}(\theta) > \lambda} \pi(\theta) \, d\theta,
\end{equation}

\noindent i.e., the cumulative prior volume where the likelihood exceeds some threshold $\lambda$. Here $\theta$ represents model parameters such as \textit{p}, $\psi$, and $\omega$, $\mathcal{L}(\theta)$ is the likelihood from the photoelectron azimuthal angle distribution, and $\pi(\theta)$ is the prior.

The Bayesian evidence quantifying how well a given model explains the data can then be written as

\begin{equation}
Z = \int_0^1 \mathcal{L}(X) \, dX.
\end{equation}

In QUEEN-BEE, the Bayesian nested sampler starts with a population of ``live points'' sampled from a prior distribution defining the parameter space over which the nested sampler explores possible models. The algorithm iteratively replaces the lowest likelihood point among a set of live points with a new one drawn from the constrained prior, shrinking the prior mass until the integral converges. At each iteration, likelihoods and volumes are accumulated to estimate the evidence and generate posterior samples. The algorithm terminates when the remaining live points contribute negligibly to the total evidence. 

\subsection{Defining Bayesian Priors in QUEEN-BEE} 
\label{subsec:priors}

Motivated by physicality, for the constant EVPA model, we adopt a uniform prior, where $q, u \in [-1, 1]$, forming a uniform circular prior volume corresponding to the PD $p = \sqrt{q^2+u^2}\leq1$. 

For the unpolarized model, a narrow Gaussian prior is defined for both $q$ and $u$ in lieu of a true delta function for consistency of evidence calculation and comparison. The Gaussian for each Stokes parameter is defined with a center at $q=u=0$ and $\sigma=10^{-6}$.

Finally, for the rotating EVPA model, the prior is defined over the parameters $q$, $u$, and $\omega$ (the angular EVPA rotation rate in degrees per day), where $q \in [-1, 1]$, $u \in [-1, 1]$, and $\omega \in [-200^\circ, 200^\circ]$. Although polarization rotation may exceed a rotation rate of $200^\circ$ per day, this prior is chosen to capture realistic EVPA evolution scenarios without introducing degeneracy or artifacts from undersampled high-frequency rotations.

\subsection{QUEEN-BEE Model Comparison}
\label{sec:model compare}

Model comparison is performed using the Bayesian evidence values $Z$ computed during nested sampling. For each model---constant EVPA, rotating EVPA, and unpolarized---the evidence quantifies the integrated likelihood over the full prior space providing a standardized metric for comparing models of different complexities. The relative support for any two models labeled 0 and 1 is expressed through the Bayes factor 

\begin{equation}
B_{10} = \frac{Z_1}{Z_0}, \quad \text{or} \quad \ln B_{10} = \ln Z_1 - \ln Z_0,
\end{equation}

\noindent where $Z_1$ and $Z_0$ refer to the evidences for each respective model. In this work, we interpret the Bayes factor using the modified Jeffreys scale \citep{Jeffreys1986}, where $B_{10} \textgreater 150$ (equivalently $lnB_{10}=\Delta  lnZ_{10} \textgreater 5$) is considered decisive evidence in favor of a model. Comparisons are made pairwise across the three hypotheses to assess whether the data prefer a rotating EVPA model over a constant one, and whether either polarized model is favored over the unpolarized null model. This approach provides a quantitative, prior-weighted assessment of model plausibility. 

\subsection{QUEEN-BEE Implementation Architecture and Usage}

QUEEN-BEE is a polarization analysis framework implemented in a modular 
Python environment structured around the Bayesian inference library \texttt{bilby} \citep{2019ApJS..241...27A}. The workflow is managed within a unified python script enabling consistent data handling and reproducible comparisons across competing EVPA models. The architecture supports systematic testing of constant, rotating, and unpolarized hypotheses using a common interface for model specification and evidence comparison. 

Priors for all model parameters are defined using \texttt{bilby.PriorDict} following the specifications outlined in section~\ref{subsec:priors}. Likelihoods are defined as described in section~\ref{subsec:models}, with function \texttt{log\_likelihood\_qu()} describing a constant EVPA and \texttt{log\_likelihood\_rotrate()} describing a constant time-evolving EVPA. All likelihood functions are compatible with \texttt{bilby}'s environment and return log-likelihoods.\footnote{We use log-likelihoods for numerical stability, as they avoid underflow from multiplying many small probabilities.}

Posterior sampling and evidence computation are performed using the \texttt{dynesty} nested sampler \citep{2020MNRAS.493.3132S} accessed through the \texttt{bilby.run\_sampler()} interface. The configuration includes adjustable parameters \texttt{nlive} (number of live points, usually set to 100) and \texttt{walks} (number of steps per proposal, usually set to 20) to influence the resolution of the posterior and robustness of the evidence estimation.

Parallelization is enabled through Python's \texttt{multiprocessing.Pool()} backend to improve efficiency for multi-dimensional models. Nested sampling produces both posterior samples for each model and estimates of the Bayesian evidence of each model for comparison.

The general workflow for QUEEN-BEE follows the same preprocessing steps as those employed in PCUBE-based analyses in \texttt{ixpeobssim}. Source photons are extracted from a circular region using \texttt{SAOImage DS9}, centered by using the inbuilt ``centroid'' feature. A source count rate is evaluated and, if necessary, background rejection is conducted following \cite{2023AJ....165..143D}. Background subtraction is handled natively within QUEEN-BEE through the selection of an annular background region of inner and outer radii, in units of arcseconds, specified with variables \texttt{r\_innerbg} and \texttt{r\_outerbg} respectively. Event selection including energy and time-binning may be performed either using \texttt{xpselect} in \texttt{ixpeobssim} before utilizing QUEEN-BEE, or natively specified within QUEEN-BEE.

Within QUEEN-BEE, the source position is set using the CCD pixel coordinates \texttt{x\_center} and \texttt{y\_center}. The circular extraction region radius is defined with the \texttt{reg} variable, given in arcseconds. Input data are provided through the \texttt{eventdir} path, while response files are specified via \texttt{modfact\_directory}, with the appropriate time-dependent response file identified by \texttt{modfact}. Output directories are named by supplying the source name to the \texttt{source} variable. Energy filtering is controlled by \texttt{emin} and \texttt{emax}, which define the lower and upper energy bounds in keV.

Each model is tested running the respective model function---\texttt{scout\_const()} for a constant EVPA model, \texttt{scout\_rot()} for a rotating model, and \texttt{scout\_unpolar()} for an unpolarized model. The sampler continues until the estimated contribution to the total evidence from the remaining live points becomes smaller than a threshold $\Delta ln(Z)$ usually set to $\Delta ln(Z)\textless 0.1$.

Each model test produces posterior distributions, marginal and joint parameter plots, best estimates of $q$, $u$, and $\omega$ values with 90\% credible intervals, and a $ln(Z)$ for each model. Bayes factors may be calculated to compare pairs of models, following the formalism outlined in section~\ref{sec:model compare}.

\section{Revisiting \ixpe Observations of GX~13$+$1 with QUEEN-BEE} \label{sec:GX~13$+$1}

\subsection{Previous Polarimetric Insights on GX~13$+$1}

GX~13$+$1 is a uniquely insightful \ixpe target due to its brightness, persistent accretion, and complex emission geometry. It has been observed by \ixpe during three separate epochs between 2023 October and 2024 April, with a total exposure of approximately 300~ks across all observations. The individual observation dates, observation IDs (ObsIDs), and detector exposures of each observation are summarized in Table~\ref{table:obsdates}. All observations were initially processed similarly, using a circular region ranging from 80-100\arcsec\ centered at the source, no background rejection or subtraction \citep{2023AJ....165..143D}, and polarimetric analysis performed using the PCUBE algorithm of \texttt{ixpeobssim} \citep{2024A&A...688A.217B, 2024A&A...688A.170B, 2025ApJ...979L..47D, 2025arXiv250805763K}.

\begin{table*}
\centering
\caption{Overview of three \ixpe observations of GX~13$+$1.}
\label{table:obsdates}
\begin{tabular}{ccccc}
            \hline\hline
 Observation & Dates & ObsID & Instrument & Exposure (s) \\
\hline
1 & 2023 Oct 17--19 & 02006801 & DU1 & 98204     \\
& &  & DU2 & 98299 \\
&  &  & DU3 & 98299 \\
2 & 2024 Feb 25--27 & 03001101 & DU1 & 90670 \\
& &  & DU2 & 90819 \\
&  &  & DU3 & 90799 \\
3 & 2024 Apr 20--23 & 03003401 & DU1 & 99886 \\
& &  & DU2 & 100011 \\
&  &  & DU3 & 100112 \\
\hline
\end{tabular}
\end{table*}

Across the three \ixpe observations of GX~13$+$1, the polarization behavior revealed a complex and variable picture. The first observation in October 2023 exhibited signficant intra-observation EVPA evolution, with a smooth rotation of approximately 70 degrees across the two-day observation \citep{2024A&A...688A.170B}. In contrast, the February 2024 observation showed a stable polarization signal and no statistically significant EVPA variability \citep{2024A&A...688A.217B}. The April 2024 observation revealed a pair of pronounced dips in the light curve, during which the spectrum hardened and PD increased accompanied by EVPA swings comparable in amplitude to those seen in the first October observation of the source \citep{2025ApJ...979L..47D}. Taken together, these results highlight that GX~13$+$1 exhibits polarization behavior that varies both between and within observations, motivating the use of time-resolved, model comparative methods such as those enabled by QUEEN-BEE.

\subsection{Data Reduction Methods}

In the following, we have re-analyzed the \ixpe observations reported in  \cite{2024A&A...688A.217B, 2024A&A...688A.170B, 2025ApJ...979L..47D}; and \cite{2025arXiv250805763K}, with the model-independent PCUBE algorithm. We use a circular source region of 80\arcsec\ for the first observation, and 100\arcsec\ for the other two observations. We do not perform background rejection for the first observation whereas we do in the other two observations, following the prescription in \cite{2023AJ....165..143D} (${\rm\,count\ rate/s/arcmin^2}$ = 2.22, 1.15, and 1.37 ${\rm\,counts/s/arcmin^2}$ respectively). We use time-dependent response versions 20230702 for the first observation and 20240101 for the other two observations. 

We then apply our Bayesian framework, QUEEN-BEE (introduced in Section \ref{sec:Methodology}), to the observations, testing three EVPA models---constant EVPA, linearly rotating EVPA, and unpolarized---using the QUEEN-BEE functions \texttt{scout\_const()}, \texttt{scout\_rot()}, and \texttt{scout\_unpolar()} respectively. A summary of the resulting $q, u,$ and $\omega$ and $ln(Z)$ obtained for each model is outlined in Table~\ref{table:all} and a model comparison overview is provided in Table~\ref{table:bfall}.

\subsection{Polarimetric Results}

In the PCUBE analysis of the three previous \ixpe observations of GX~13$+$1 in the 2--8~keV band, we find:\footnote{All uncertainties reported for PCUBE results are $1\sigma$} the first observation yields a PD of $1.2\pm0.3\%$ with a polarization angle (PA) of $1\pm7^{\circ}$ at a significance of $3.6\sigma$; the second observation shows a higher PD of $2.5\pm0.3\%$ with PA $25\pm4^{\circ}$ at $6.9\sigma$; and the third observation returns a PD of $1.2\pm0.3\%$ with PA $-12\pm7^{\circ}$ at $3.6\sigma$.

We additionally report energy-binned PCUBE analysis for the first \ixpe observation of GX~13$+$1 for comparison with subsequent QUEEN-BEE results. In 2--4~keV we obtain a PD of $1.4 \pm 0.3 \%$ and a PA of $16\pm6^\circ$ at a $4.2\sigma$ confidence level, and in the 4--8~keV band we obtain a PD of $1.5 \pm 0.5 \%$ and a PA of $-17\pm9^\circ$ at a $2.6\sigma$ confidence level. 

We lastly perform time-binned PCUBE analysis following the binning performed in \cite{2025ApJ...979L..47D}, using six time bins corresponding to before, during, and after two dips in the light curve. In the third time bin (immediately after the first light curve dip and between 15.9 hours and 27.0 hours after the start of the observation), we obtain a PD of $1.4 \pm 0.6 \%$ and a PA of $2\pm18^\circ$ at a $1.0\sigma$ confidence level. 

We perform an analogous 2--8~keV analysis with QUEEN-BEE to enable direct comparison with PCUBE (Table \ref{table:all}). We derive Bayesian evidences ($lnZ$) for unpolarized, constant EVPA, and rotating EVPA models and calculate Bayes factors from these, which we interpret with the modified Jeffreys scale (Table \ref{table:bfall}).

\noindent \textbf{Observation~1}: Bayes factors indicate decisive preference for the rotating EVPA model over both constant and unpolarized models; the rotating EVPA fit yields a PD of \(2.6^{+0.3}_{-0.2}\%\), a PA of $177\pm3^\circ$, and an EVPA rotation rate of $42\pm4\,^\circ/\mathrm{day}$.

\noindent \textbf{Observation~2}: Bayes factors indicate decisive preference for a constant EVPA model over an unpolarized one and marginal preference over a rotating EVPA model; the constant fit yields a PD of \(2.5\pm0.5\%\) and a PA of $27^{+5}_{-6}\,^\circ$.

\begin{figure}[ht]
    \centering
    \includegraphics[width=0.45\textwidth, trim=0 20 0 40, clip]{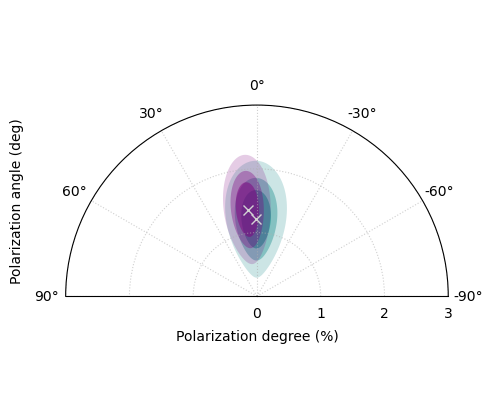}
    \vspace{-0.7em}
    
    \includegraphics[width=0.45\textwidth, trim=0 10 0 40, clip]{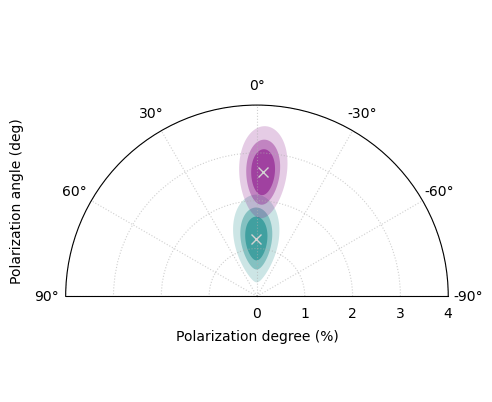}
    \caption{
    Polarimetric summary for the first GX~13$+$1 \ixpe observation, with EVPA plotted east of north.
    \textbf{Top:} Comparison between PCUBE (teal) and QUEEN-BEE assuming constant EVPA (\texttt{scout\_const()}) (purple).
    \textbf{Bottom:} Comparison between PCUBE (teal) and QUEEN-BEE assuming rotating EVPA (\texttt{scout\_rot()}) (purple).
    }
    \label{fig:gx13o1}
\end{figure}

\begin{figure}
    \centering
    \includegraphics[width=0.45\textwidth, trim=0 20 0 40, clip]{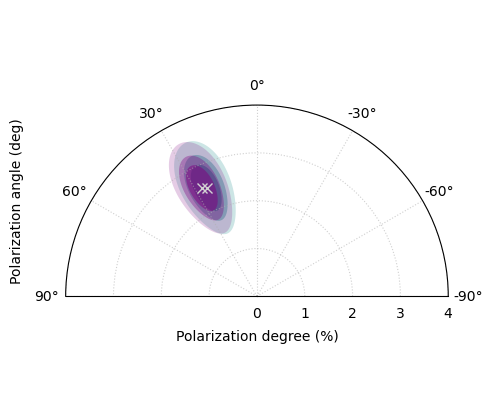}
    \vspace{-0.7em}
    \caption{
    Polarimetric summary for the second GX~13$+$1 \ixpe observation, with EVPA plotted east of north. Comparison between PCUBE (teal) and QUEEN-BEE assuming constant EVPA (\texttt{scout\_const()}) (purple).}
    \label{fig:gx13o2}
\end{figure}

\begin{figure}
    \centering
    \includegraphics[width=0.45\textwidth, trim=0 20 0 40, clip]{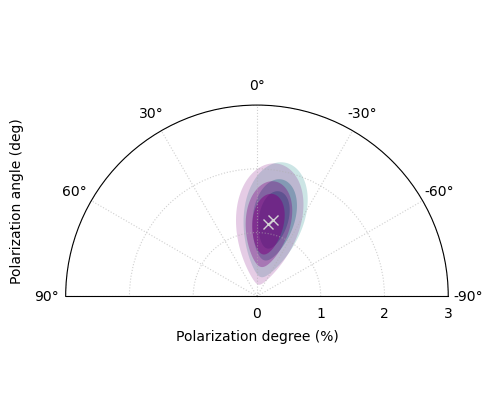}
    \vspace{-0.7em}
    
    \includegraphics[width=0.45\textwidth, trim=0 10 0 40, clip]{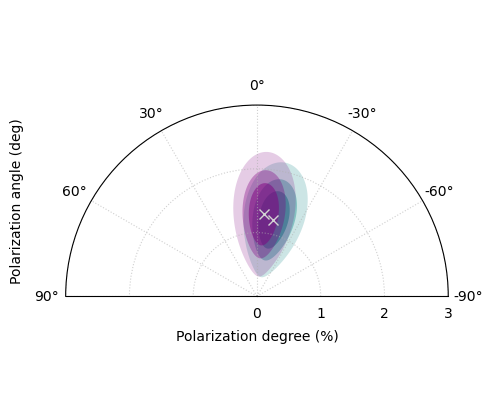}
    \caption{
    Polarimetric summary for the third GX~13$+$1 \ixpe observation, with EVPA plotted east of north.
    \textbf{Top:} Comparison between PCUBE (teal) and QUEEN-BEE assuming constant EVPA (\texttt{scout\_const()}) (purple).
    \textbf{Bottom:} Comparison between PCUBE (teal) and QUEEN-BEE assuming rotating EVPA (\texttt{scout\_rot()}) (purple).
    }
    \label{fig:gx13o3}
\end{figure}

\noindent \textbf{Observation~3}: Bayes factors indicate decisive preference for a rotating EVPA model over an unpolarized model and weak preference over a constant EVPA model; the rotating fit yields a PD of \(1.3\pm0.2\%\), a PA of $-5^{+7}_{-5}\,^\circ$, and an EVPA rotation rate of $20\pm9\,^\circ/\mathrm{day}$; the constant fit yields a PD of \(1.14^{+0.26}_{-0.12}\%\) and a PA of $171^{+8}_{-5}\,^\circ$.
Full-band (2--8~keV) QUEEN-BEE and PCUBE results are jointly plotted in Figs.~\ref{fig:gx13o1}, \ref{fig:gx13o2}, and \ref{fig:gx13o3}.

\begin{figure}
    \centering
    \includegraphics[width=0.45\textwidth, trim=0 20 0 40, clip]{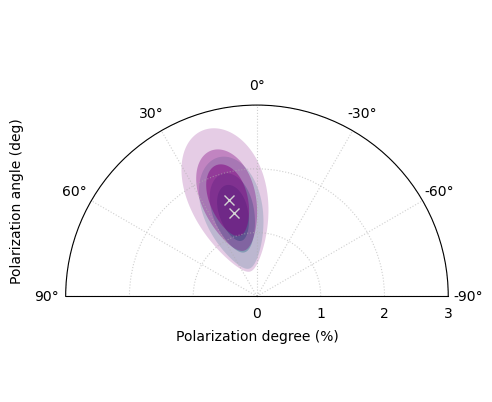}
    \vspace{-0.7em}
    
    \includegraphics[width=0.45\textwidth, trim=0 10 0 40, clip]{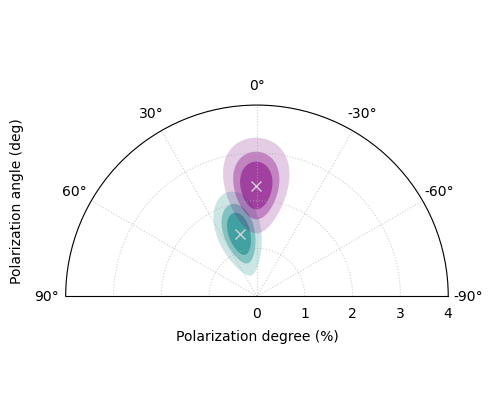}
    \caption{
    Polarimetric summary for the first GX~13$+$1 \ixpe observation in the 2--4~keV energy band, with EVPA plotted east of north.
    \textbf{Top:} Comparison between PCUBE (teal) and QUEEN-BEE assuming constant EVPA (\texttt{scout\_const()}) (purple).
    \textbf{Bottom:} Comparison between PCUBE (teal) and QUEEN-BEE assuming rotating EVPA (\texttt{scout\_rot()}) (purple).
    }
    \label{fig:gx13o1 2-4}
\end{figure}

\begin{figure}
    \centering
    \includegraphics[width=0.45\textwidth, trim=0 20 0 40, clip]{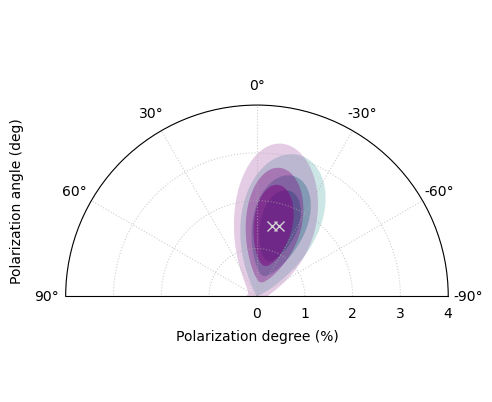}
    \vspace{-0.7em}
    
    \includegraphics[width=0.45\textwidth, trim=0 10 0 40, clip]{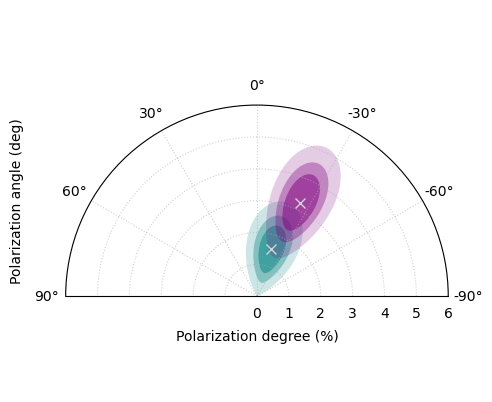}
    \caption{
    Polarimetric summary for the first GX~13$+$1 \ixpe observation in the 4--8~keV energy band, with EVPA plotted east of north.
    \textbf{Top:} Comparison between PCUBE (teal) and QUEEN-BEE assuming constant EVPA (\texttt{scout\_const()}) (purple).
    \textbf{Bottom:} Comparison between PCUBE (teal) and QUEEN-BEE assuming rotating EVPA (\texttt{scout\_rot()}) (purple).
    }
    \label{fig:gx13o1 4-8}
\end{figure}

Energy-resolved analysis of the first \ixpe observation of GX~13$+$1 (2--4 and 4--8~keV) produce Bayes factors decisively favoring the EVPA rotation model in both bands---\textbf{2--4~keV:} PD $= 2.3^{+0.4}_{-0.3}\%$, PA $= 0^{+6}_{-4}\,^\circ$, EVPA rotation rate $= 37\pm6\,^\circ/\mathrm{day}$, and
\textbf{4--8~keV:} PD $= 3.2\pm{0.4}\%$, PA $= -6^{+4}_{-3}\,^\circ$, EVPA rotation rate $= 48^{+6}_{-7}\,^\circ/\mathrm{day}$.
Numerical summaries and Bayes factors are reported in Tables~\ref{table:energy} and \ref{table:bf_energy_split}; plots in Figs.~\ref{fig:gx13o1 2-4} and \ref{fig:gx13o1 4-8}.

\begin{figure}
    \centering
    \includegraphics[width=0.45\textwidth, trim=0 20 0 40, clip]{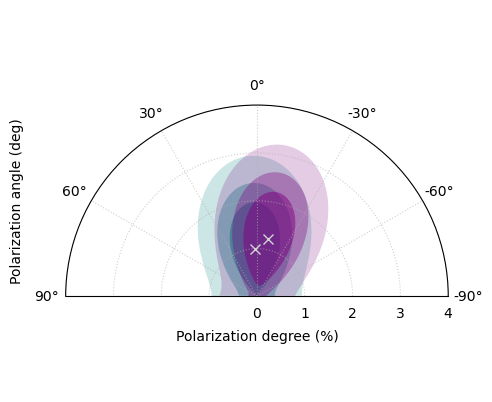}
    \vspace{-0.7em}
    
    \includegraphics[width=0.45\textwidth, trim=0 10 0 40, clip]{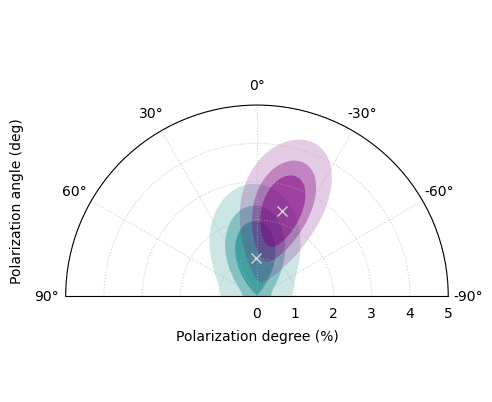}
    \caption{
    Polarimetric summary for the third GX~13$+$1 IXPE observation between 15.9 hrs-27.0 hrs, with EVPA plotted east of north.
    \textbf{Top:} Comparison between PCUBE (teal) and QUEEN-BEE assuming constant EVPA (\texttt{scout\_const()}) (purple).
    \textbf{Bottom:} Comparison between PCUBE (teal) and QUEEN-BEE assuming rotating EVPA (\texttt{scout\_rot()}) (purple).
    }
    \label{fig:gx13o3 minibin}
\end{figure}

Lastly, for the QUEEN-BEE time-resolved analysis of Observation~3, the third post-dip bin (15.9--27.0~h after start) decisively favors EVPA rotation, with a PD of \(2.32^{+0.29}_{-0.12}\%\), a PA of $-164^{+6}_{-2}\,^\circ$, and an EVPA rotation rate of $170^{+20}_{-40}\,^\circ/\mathrm{day}$ (Tables~\ref{table:obs3small}, \ref{table:bf3small}; Fig.~\ref{fig:gx13o3 minibin}).

\begin{table*}
\centering
\caption{QUEEN-BEE results for three previous \ixpe observations of GX~13$+$1 in 2--8~keV. Uncertainties are reported as $90\%$ credible intervals.}
\label{table:all}
\begin{tabular}{cccccc}
\hline\hline
\textbf{Observation} & \textbf{Model} & $\mathbf{q}$ & $\mathbf{u}$ & $\mathbf{\omega(^\circ/day)}$ & $\mathbf{ln(Z)}$\\
\hline
1 & Unpolarized & $0\pm0$ & $0\pm0$ & - & $76762640.5 \pm 0.3$ \\
& Constant EVPA & $0.013\pm0.003$ & $0.003\pm0.002$ & - & $76762659.6 \pm  0.8$ \\
& Rotating EVPA & $0.026\pm0.003$ & $-0.003\pm0.003$ & $42\pm4$ & $76762736.1 \pm 0.2$ \\
\hline
2 & Unpolarized & $0\pm0$ & $0\pm0$ & - & $54215304.1\pm0.3$ \\
& Constant EVPA & $0.015\pm0.003$ & $0.020^{+0.004}_{-0.003}$ & - & $54215368.3\pm0.4$ \\
& Rotating EVPA & $0.016\pm0.003$ & 
$-0.020^{+0.003}_{-0.004}$ & $-8^{+7}_{-6}$ & $54215366.4\pm0.1$ \\
\hline
3 & Unpolarized & $0\pm0$ & $0\pm0$ & - & $71317165.5\pm0.3$ \\
& Constant EVPA & $0.011^{+0.003}_{-0.003}$ & $-0.003\pm0.003$ & - & $71317173.4\pm0.7$ \\
& Rotating EVPA & $0.013\pm0.003$ & 
$-0.002\pm0.003$ & $19^{+9}_{-8}$ & $ 71317174.8\pm0.5$ \\
\hline
\end{tabular}
\end{table*}

\begin{table*}
\centering
\caption{Log Bayes factors ($\ln B$) and polarization model comparison for three previous \ixpe observations of GX~13$+$1 in the full 2--8~keV band with propagated 1$\sigma$ uncertainties.}
\label{table:bfall}
\begin{tabular}{ccc}
\hline\hline
\textbf{Observation} & \textbf{Model Comparison} & $\mathbf{\ln B_{10}}$\\
\hline
1 & Rotating vs Constant EVPA & $76.5 \pm 0.8$\\
& Constant EVPA vs Unpolarized & $19.1 \pm 0.8$\\
& Rotating vs Unpolarized      & $95.6 \pm 0.3$\\
\hline
2 & Rotating vs Constant EVPA & $-1.9 \pm 0.4 $\\
& Constant EVPA vs Unpolarized & $64.2 \pm 0.5$\\
& Rotating vs Unpolarized      & $62.3 \pm 0.3$\\
\hline
3 & Rotating vs Constant EVPA & $1.4 \pm 0.8 $\\
& Constant EVPA vs Unpolarized & $7.9 \pm 0.8$\\
& Rotating vs Unpolarized & $9.2 \pm 0.5$\\
\hline
\end{tabular}
\end{table*}

\begin{table*}
\centering
\caption{QUEEN-BEE results for the first \ixpe observation of GX~13$+$1 (ObsID 02006801) split into 2--4~keV and 4--8~keV energy bands. Uncertainties are reported as $90\%$ credible intervals.}
\label{table:energy}
\begin{tabular}{lcccc}
\hline
\hline
\textbf{Model} & $\mathbf{q}$ & $\mathbf{u}$ & $\mathbf{\omega(^\circ/day)}$ & $\mathbf{ln(Z)}$\\
\hline
\multicolumn{5}{l}{\textbf{2--4~keV}} \\
Unpolarized & $0\pm0$ & $0\pm0$ & - & $62111305.9\pm0.3$\\
Constant EVPA & $0.013\pm0.004$ & $0.008\pm0.004$ & - & $62111318.9\pm0.4$ \\
Rotating EVPA & $0.023\pm0.003$ & $0.00027^{+0.0040}_{-0.0040}$ & $37\pm6$ & $62111348.2\pm0.1$\\
\hline
\multicolumn{5}{l}{\textbf{4--8~keV}} \\
Unpolarized & $0\pm0$ & $0\pm0$ & - & $12595987.8\pm0.1$ \\
Constant EVPA & $0.014^{+ 0.006}_{-0.005}$ & $-0.006\pm0.005$ & - & $12595990.9\pm0.4$\\
Rotating EVPA & $ 0.032^{+0.004}_{-0.005}$ & $-0.006\pm0.005$ & $48^{+6}_{-7}$ & $12596034.6\pm0.4$ \\
\hline
\end{tabular}
\end{table*}

\begin{table*}[htbp]
\centering
\caption{Model comparison for the first \ixpe observation of GX~13+1 (ObsID 02006801), split into 2--4~keV and 4--8~keV energy bands. Log Bayes factors ($\ln B$) are shown with 1$\sigma$ propagated uncertainties.}
\label{table:bf_energy_split}
\begin{tabular}{lc}
\hline\hline
\textbf{Model Comparison} & $\mathbf{\ln B_{10}}$\\
\hline
\multicolumn{2}{l}{\textbf{2--4~keV}} \\
Rotating vs Constant EVPA & $29.3 \pm 0.4$\\
Constant EVPA vs Unpolarized & $13.0 \pm 0.5$\\
Rotating vs Unpolarized & $42.2 \pm 0.3$\\
\hline
\multicolumn{2}{l}{\textbf{4--8~keV}} \\
Rotating vs Constant EVPA    & $43.8 \pm 0.6$\\
Constant EVPA vs Unpolarized & $3.1 \pm 0.4$\\
Rotating vs Unpolarized & $46.9 \pm 0.4$\\
\hline
\end{tabular}
\end{table*}

\begin{table*}
\centering
\caption{QUEEN-BEE Results for the third \ixpe observation of GX~13$+$1 (ObsID 03003401) for 2--8~keV between 15.9 hrs - 27.0 hrs. Uncertainties are reported as $90\%$ credible intervals}
\label{table:obs3small}
\begin{tabular}{lcccc}
            \hline\hline
     \textbf{Model} & $\mathbf{q}$ & $\mathbf{u}$ & $\mathbf{\omega(^\circ/day)}$ & $\mathbf{ln(Z)}$\\
\hline
            Unpolarized & $0\pm0$ & $0\pm0$ & - & $14254086.8\pm0.1$ \\
            Constant EVPA & $0.011^{+0.006}_{-0.007}$ & $-0.004^{+0.006}_{-0.007}$ & - & $14254081.5\pm0.2$ \\
            Rotating EVPA & $0.020\pm0.006$ & 
 $-0.012^{+0.007}_{-0.006}$ & $170^{+20}_{-40}$ & $ 14254093.0\pm0.4$ \\
            \hline
         \end{tabular}
   \end{table*}

\begin{table*}
\centering
\caption{Log Bayes factors ($\ln B$) and polarization model comparison for the third \ixpe observation of GX~13$+$1 (ObsID 03003401) between 15.9 hrs - 27.0 hrs with propagated 1$\sigma$ uncertainties.}
\label{table:bf3small}
\begin{tabular}{lc}
\hline\hline
\textbf{Model Comparison} & $\mathbf{\ln B_{10}}$\\
\hline
Rotating vs Constant EVPA & $11.4 \pm 0.4 $\\
Constant EVPA vs Unpolarized & $-5.2 \pm 0.2$\\
Rotating vs Unpolarized & $6.2 \pm 0.4$\\
\hline
\end{tabular}
\end{table*}

\section{Discussion}
\label{sec:discussion}

The three IXPE observations of GX~13$+$1 analyzed in this work reveal a complex and dynamic picture of the system's X-ray polarization behavior. Using the QUEEN-BEE framework, we quantitatively evaluate the evidence for constant, rotating, and unpolarized EVPA models in a statistically rigorous manner, revealing both temporal and spectral variability that is not easily recovered through traditional time-binned methods. In instances where the preferred QUEEN-BEE model is one of constant polarization---such as the full 2--8~keV results for the second and third \ixpe observations of GX~13$+$1---results between PCUBE and QUEEN-BEE agree very well (see Figures \ref{fig:gx13o2} and \ref{fig:gx13o3}.) We additionally highlight instances of QUEEN-BEE results providing more detailed insight into the source's behavior.

The first \ixpe observation exhibits decisive evidence for smooth EVPA rotation, with a rotation rate of $\sim$42$^\circ/day$ and consistently high Bayes factors in favor of the rotating model across both full and energy-resolved bands. These results are consistent with PCUBE analyses that suggested an EVPA rotation of $\sim$70$^\circ$ in two days \citep{2024A&A...688A.170B}. The results also illustrates a trend of marginal increase in PD from $1.2\pm0.3\%$ using PCUBE to a PD of $2.6^{+0.3}_{-0.2}\%$ using QUEEN-BEE as shown in Figure~\ref{fig:gx13o1}. Notably, the constant EVPA model of QUEEN-BEE provides statistically consistent results to the PCUBE results of this observation.

The second \ixpe observation shows a notably different polarization behavior. The QUEEN-BEE analysis reveals a stable EVPA across the observation, with the constant EVPA model decisively preferred over unpolarized models and weakly preferred to the rotating model. This observation yields the highest PD among all time- and energy-averaged epochs, suggesting in the absence of EVPA rotation or depolarizing variability, the intrinsic polarization signal is more fully recovered. This contrast with the first \ixpe observation suggests that geometric or dynamical changes in the system, potentially linked to orbital phase or viewing angle, play a significant role in shaping the observed polarization behavior. The QUEEN-BEE and PCUBE observations are in strong agreement, as illustrated by Figure~\ref{fig:gx13o2}, following the pattern of agreeing results between PCUBE and the constant EVPA model of QUEEN-BEE, thus verifying the veracity of our QUEEN-BEE framework.

\begin{figure}
    \centering
    \includegraphics[width=0.45\textwidth]{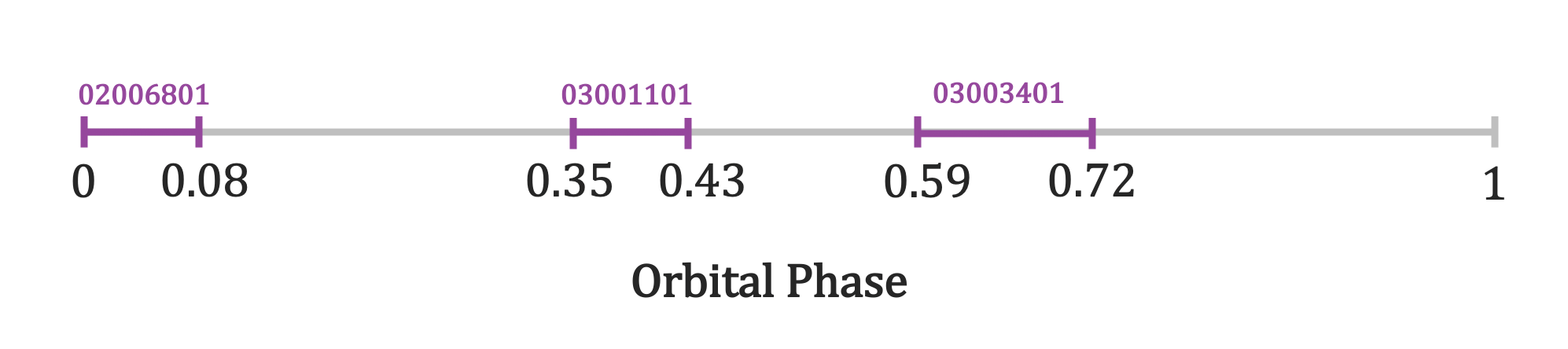}
    \vspace{-0.7em}
    \caption{
    Schematic illustration of the orbital phase intervals within GX~13$+$1’s 24.5-day period sampled by the three previous \ixpe observations. Phase 0 is defined as the start of the first \ixpe observation of the source, defined as MJD 60234.718.}
    \label{fig:orbitalphase}
\end{figure}

The third \ixpe observation presents a hybrid case, with QUEEN-BEE favoring the rotating EVPA model weakly over the constant model when analyzing the full observation (see Figure~\ref{fig:gx13o3}), but revealing decisive evidence for EVPA rotation in time-resolved analysis focused on the post-dip segment between 15.9 and 27.0 hours (see Figure~\ref{fig:gx13o3 minibin}. In the time-averaged analysis, both the rotating and constant EVPA models are strongly favored over the unpolarized model with both model results agreeing with the PCUBE result.

The QUEEN-BEE framework provides novel insight into the polarimetric properties of GX~13$+$1 in both time- and energy- binned analysis of data where frequentist analysis ignores rotational behavior of the EVPA resulting in sub-threshold significance. These new constraints provide interesting physical implications for the source.

In the first \ixpe observation, energy-binned analysis using PCUBE of even two energy bins drops the polarimetric results below $3\sigma$ confidence. Specifically, in the 4--8~keV band the PCUBE results are reported with $2.6\sigma$ confidence. The QUEEN-BEE result shows decisive evidence in favor of a rotating EVPA model over both constant EVPA and unpolarized models with marginal increases in the reported PDs in both energy bins reflecting the expected rise in PD when accounting for EVPA rotation cancelling polarization signal. This marginal difference in PD has dramatic implications for the expected coronal geometry of GX~13$+$1. 

X-ray spectroscopic emission of NS-LMXBs of both Z-state and atoll-state sub-classes is traditionally modeled with a soft thermal emission (blackbody) component corresponding with the NS surface and a hard Comptonized region corresponding with the NS coronal plasma. Two prevalent models exist to describe the physics of the system. The ``eastern'' model predicts a colder disk blackbody component and hotter Comptonized plasma component, while the ``western'' model suggests a hot NS blackbody and cold corona-Comptonized disk \citep{1984PASJ...36..741M, 1988ApJ...324..363W, 1996PASJ...48..257A}. These two models contain significant spectroscopic degeneracy, though X-ray polarimetric measurements have long thought to be a promising avenue for constraining models \citep{2022MNRAS.514.2561G}. 

Various coronal geometries have been predicted by the eastern and western spectral models. In particular, the slab geometry (a flat corona extended over the surface of the accretion disk) is associated with the western model. In this picture, the disk photons of an extended coronal slab would experience high scattering and the NS blackbody radiation would be directly detectable at low inclinations. This combination produces the predicted hot NS blackbody and cold disk of the western model \citep{1994ApJ...434..570T, 1996ApJ...470..249P, 2019ApJ...875..148Z}. The shell geometry, in contrast (a spherical shell surrounding the NS) favors the eastern model as the NS photons get Comptonized by the surrounding corona while leaving the disk unobstructed by any extended coronal emission \citep{2019ApJ...875..148Z, 2022MNRAS.514.2561G}. 

\begin{figure}
    \centering
    \includegraphics[width=0.45\textwidth, trim=0 2 0 7, clip]{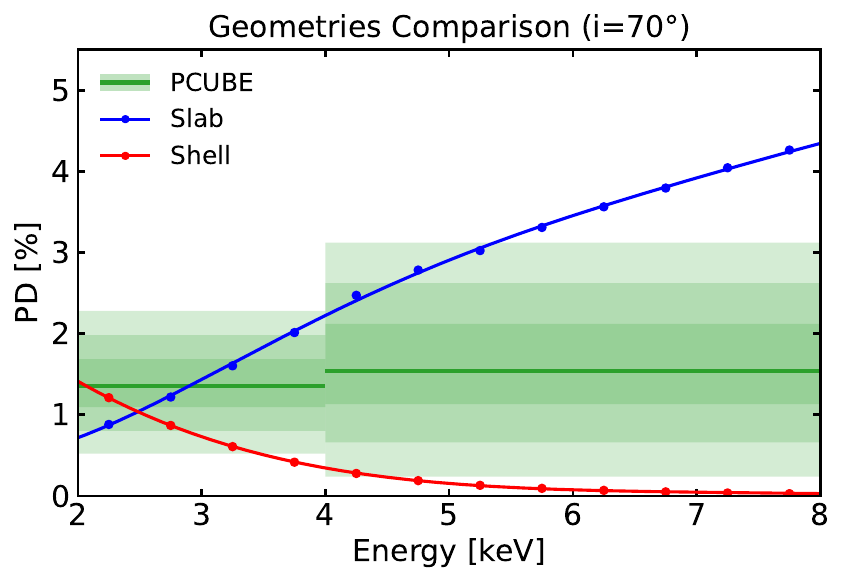}
    \vspace{0.5em}
    
    \includegraphics[width=0.45\textwidth, trim=0 2 0 7, clip]{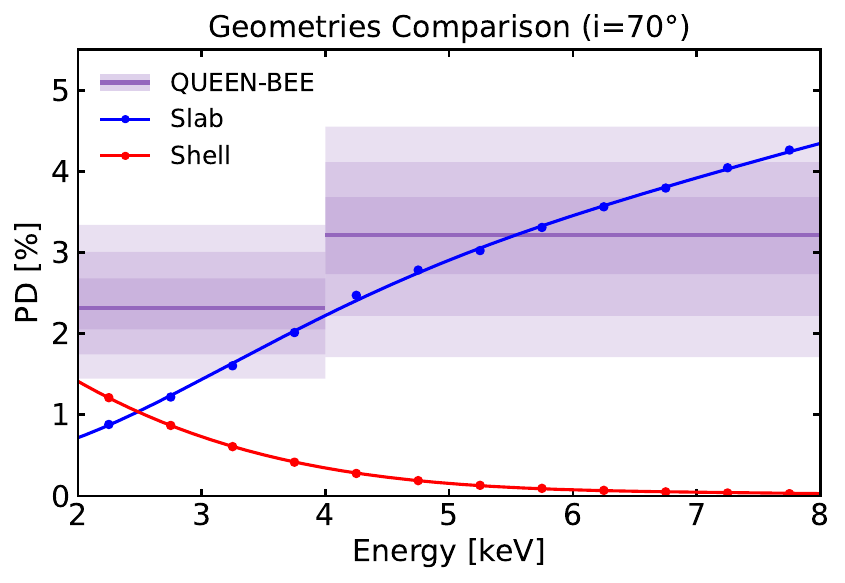}
    \caption{
    Polarimetric summary for the first GX~13$+$1 \ixpe observation in the 2--4 and 4--8~keV energy bands overplotted with MONK simulations of coronal geometries from \cite{2022MNRAS.514.2561G}.
    \textbf{Top:} Comparison between PCUBE (green) and MONK simulations of slab and shell corona. The 2--4~keV result has a $4.1\sigma$ significance, while 4--8~keV has a $2.6\sigma$ significance.
    \textbf{Bottom:} Comparison between QUEEN-BEE (purple) assuming rotating EVPA and MONK simulations of slab and shell corona.}
    \label{fig:gx13p1corona}
\end{figure}

Relativistic Monte Carlo simulations of X-ray polarization performed with the MONK code \citep{2019ApJ...875..148Z, 2022MNRAS.515.2882Z} and adapted in \cite{2022MNRAS.514.2561G} to describe (among other scenarios) weakly magnetized NS-LMXBs in the high soft state at inclination of $70^\circ$---descriptive of GX~13$+$1 \citep{2012A&A...543A..50D, 2024MNRAS.52711855G}---indicate that a shell geometry would result in a negative trend of PD as a function of increasing energy across the 2--8~keV band, while a slab geometry would result in a rise of PD across the 2--8~keV band. The PCUBE results fail to exclude either geometry in energy-resolved analysis due to large uncertainties associated with polarimetric results, though the 4--8~keV band polarization falling below MDP99 (1.54\% PD) appear to slightly prefer a shell geometry (see Figure \ref{fig:gx13p1corona}). QUEEN-BEE energy-resolved analysis, however, appears to exclude the shell geometry favoring a slab corona instead, as shown in Figure \ref{fig:gx13p1corona}. These results not only highlight the favorability of a western model in describing the physical environment of GX~13$+$1, but also illustrate how even marginal refinements in the measured PD can change the interpretation of source properties like coronal geometry.

In the time-binned analysis, the recovered EVPA rotation rate after the first dip in the third \ixpe observation reaches $\sim$174$^\circ/\mathrm{day}$, more than four times the rate observed in the first \ixpe observation. This ability to distinguish variability between dips across epochs is enabled through the use of QUEEN-BEE, as PCUBE analysis of this individual time bin yields a $1\sigma$ polarimetric result, requiring stacking of multiple epochs of data to pull out EVPA rotation estimates.

This result strongly suggests that the EVPA rotation of GX~13$+$1 is a transient phenomenon, possibly linked to specific geometric changes in the accretion flow or episodic obscuration by disk winds. The alignment between the observed EVPA swing and light curve dips in the third \ixpe observation supports scenarios in which temporary absorption of the central X-ray source by clumpy or variable wind material alters the scattering geometry, producing measurable EVPA swings as suggested in \cite{2025ApJ...979L..47D}, though the source of the variability in EVPA rotation rate between epochs remains unclear. The smooth rotation of EVPA across the entire first \ixpe observation (not just associated with the observed light curve dip) additionally complicates the picture of a variable absorber exclusively causing EVPA rotation and potentially indicates a geometric effect such as spin and orbital axis misalignment as noted in \cite{2024A&A...688A.170B}.

The observed change in polarimetric behavior coincident with light curve dips in GX~13$+$1 mirrors many X-ray polarimetric properties of the dipping neutron star X-ray binary 4U~1624$-$49. In 4U~1624$-$49, significant polarization is detected in non-dip intervals (PD of $3.1 \pm 0.7\%$), while only constrained by an upper limit ($22\%$ at $99\%$ confidence) in the dip intervals. The non-dip behavior is similarly high in PD as GX~13$+$1, favoring again a slab coronal geometry \citep{2024A&A...690A.230G, 2024ApJ...963..133S}. Despite these parallels in behavior, the known presence of a persistent accretion disk wind in GX~13$+$1 from high-resolution X-ray spectroscopy \citep{2004ApJ...609..325U, 2018ApJ...861...26A, 2025ApJ...986...41R} complicates a more conclusive coronal geometry picture, as recent modeling efforts of X-ray binary wind behavior disagrees on the net effect of winds on the measured PD \citep{2024MNRAS.527.7047T, 2025A&A...694A.230N}. Analytical methods suggest the ability of disk winds to independently and significantly elevate the PD of X-ray binary systems \citep{2025A&A...694A.230N}, while Monte Carlo radiative transfer simulations of thermal-radiative disk winds suggest winds may polarize light orthogonal to the disk resulting in a slightly depolarized signal \citep{2024MNRAS.527.7047T}.

GX~13$+$1's persistent winds make it a strong candidate for performing more rigorous future investigations to resolve ambiguities in wind-related polarization behavior. Coordinated efforts between \ixpe and high-resolution X-ray spectroscopy from observatories like \xrism could aid in this process by linking the observed polarimetric behavior with spectroscopic signatures quantifying disk wind behavior such as blueshifted absorption features, hydrogen column density, and ionization levels. Moreover, such an investigation might provide a compelling conceptual analog for characterizing ``changing look'' AGN---active galaxies with spectral classifications changing on timescales of months to years, possibly due to variable obscuration from the clumpy torus or disk wind \citep{2003MNRAS.342..422M}. Radiative-transfer predictions show that variable obscuration by clumpy winds/torus might induce PA rotations similar to the potentially wind-driven changes in polarization behavior of GX~13$+$1 \citep{2013MNRAS.436.1615M}.

\section{Conclusion}
\label{sec:conclusion}
In this work, we introduced ``Q-U Event-by-Event Nested sampling for Bayesian EVPA Evolution'' (QUEEN-BEE), a Bayesian nested sampling framework designed to analyze EVPA rotation in X-ray polarimetric data. Applying QUEEN-BEE to the first three epochs of \ixpe observations of the neutron star low-mass X-ray binary GX~13$+$1, we quantitatively compared constant, rotating, and unpolarized polarization models using Bayesian evidence.

We find:
\begin{itemize}
    \item The first \ixpe observation exhibits decisive evidence for a smoothly rotating EVPA, with a rotation rate of $\sim$42$^\circ/\mathrm{day}$ and energy-dependent trends in PD and PA.
    \item The second \ixpe observation shows a stable EVPA, with a strong polarization detection but weak evidence against rotation, indicating a more symmetric or static scattering environment.
    \item The third \ixpe observation displays time-variable behavior, with evidence for localized EVPA rotation coinciding with light curve dips, supporting scenarios involving disk wind obscuration or changing accretion geometry.
    
\end{itemize}

While PCUBE and QUEEN-BEE maintain strong agreement in high-count and non-rotating EVPA regimes, in specific energy- and time-binned scenarios, QUEEN-BEE's ability to explore more complex polarization models can help address otherwise marginal results through rotation-informed polarization analysis. In the energy-resolved analysis of the first \ixpe observation, for example, QUEEN-BEE identifies decisive evidence for EVPA rotation in both the 2--4 and 4--8~keV bands while the PCUBE results in the higher-energy band fall below $3\sigma$ significance. Similarly, in the time-resolved analysis of the third \ixpe observation, QUEEN-BEE recovers a robustly rotating EVPA during a single $\sim$10-hour window coinciding with a light curve dip, despite PCUBE reporting $\sim$1$\sigma$ polarization in the same bin. These cases highlight the advantage of QUEEN-BEE in maintaining sensitivity in low-count regimes where binned methods struggle.

Beyond improved performance in low-count regimes, the QUEEN-BEE framework delivers more precise and physically informative polarization measurements by directly inferring rotation rates and uncertainties from unbinned event data. This enables detection of evolving EVPA behavior across epochs without requiring joint binning of separate observations, as required for obtaining a significant detection with PCUBE. For instance, QUEEN-BEE independently recovers a rotation rate increasing from $\sim$42$^\circ$/day in the first \ixpe observation of GX~13$+$1 to $\sim$174$^\circ$/day in the time-resolved segment of the third \ixpe observation. It also consistently yields marginally higher polarization degrees than PCUBE, particularly in energy- and time-binned analyses, by coherently modeling EVPA evolution and avoiding cancellation.

The result of these two improvements of the QUEEN-BEE analysis is additional physical insight into the dynamics of GX~13$+$1. The marginal increase in PD obtained in energy-resolved analysis shifts the data from being in agreement with shell-like coronal geometries to slab-like coronal geometries. Additionally, the discovery of an evolving EVPA rotation rate across the three \ixpe epochs indicates evidence of dynamic conditions within the disk winds, although further polarimetric studies of the source are required to fully identify the source of the dramatic change in the EVPA rotation rate seen between observations 1 and 3.

The ability to distinguish between polarization models with full statistical rigor makes QUEEN-BEE broadly applicable to a wide range of future IXPE targets, though the model-dependent nature of QUEEN-BEE means that it is best utilized in scenarios with physical motivation for smoothly rotating EVPA. Introducing additional models under the QUEEN-BEE framework which include elements like orbital- or energy-dependent rotation and multi-component rotating models, will expand the range of rotational scenarios QUEEN-BEE can evaluate. These additions, along with applying this framework to different astrophysical source classes, will be the focus of future work. As polarimetric timing emerges as a powerful diagnostic of accretion physics and coronal structure, QUEEN-BEE offers a robust and extensible framework for extracting deeper physical insights from both \ixpe and next-generation observatories such as the Rocket Experiment Demonstration of a Soft X-ray Polarimeter \citep[REDSoX;][]{2017SPIE10397E..0KM, 2025ApJ...987..113K} and the enhanced X-ray Timing and Polarimetry mission \citep[eXTP;][]{2019SCPMA..6229502Z, 2025arXiv250608101Z}.

\begin{acknowledgments}
The Imaging X-ray Polarimetry Explorer (IXPE) is a joint US and Italian mission.  The US contribution is supported by the National Aeronautics and Space Administration (NASA) and led and managed by its Marshall Space Flight Center (MSFC), with industry partner Ball Aerospace (contract NNM15AA18C).  The Italian contribution is supported by the Italian Space Agency (Agenzia Spaziale Italiana, ASI) through contract ASI-OHBI-2022-13-I.0, agreements ASI-INAF-2022-19-HH.0 and ASI-INFN-2017.13-H0, and its Space Science Data Center (SSDC) with agreements ASI-INAF-2022-14-HH.0 and ASI-INFN 2021-43-HH.0, and by the Istituto Nazionale di Astrofisica (INAF) and the Istituto Nazionale di Fisica Nucleare (INFN) in Italy.
This research used data products provided by the IXPE Team (MSFC, SSDC, INAF, and INFN) and distributed with additional software tools by the High-Energy Astrophysics Science Archive Research Center (HEASARC), at NASA Goddard Space Flight Center (GSFC).  
Funding for this work was provided in part by contract 80MSFC17C0012 from MSFC to MIT in support of the IXPE project and IXPE GO grants 80NSSC24K1175 and 80NSSC24K1748 from the NASA Goddard Space Flight Center.

During this project, S.R. received support from the Barish-Weiss Fellowship. S.R. acknowledges the valuable advice, mentorship, and support provided by Dr. Rainer Weiss. M.N. is a Fonds de Recherche du Québec – Nature et Technologies (FRQNT) postdoctoral fellow.
\end{acknowledgments}

\facilities{IXPE \citep{2022JATIS...8b6002W,2021AJ....162..208S}}

\software{astropy \citep{astropy:2013, astropy:2018, astropy:2022}, NumPy \citep{harris2020array}, SciPy \citep{2020SciPy-NMeth}, IPython \citep{PER-GRA:2007}, bilby \citep{2019ApJS..241...27A}, dynesty \citep{2020MNRAS.493.3132S}, ixpeobssim \citep{2022SoftX..1901194B}, QUEEN-BEE \citep{ravi2025queenbee}}

\bibliography{queenbee}{}
\bibliographystyle{aasjournalv7}

\end{document}